\documentclass[useAMS,usenatbib,usegraphicx]{mn2e}
\usepackage{graphicx}                                            
\usepackage{times}
\usepackage{float}
\usepackage{rotating}
\usepackage{epstopdf}
\usepackage{multirow}
\usepackage{color}

\def\ltsima{$\; \buildrel < \over \sim \;$}
\def\simlt{\lower.5ex\hbox{\ltsima}}
\def\gtsima{$\; \buildrel > \over \sim \;$}
\def\simgt{\lower.5ex\hbox{\gtsima}}
\def\gsimeq
{\hbox{\raise0.5ex\hbox{$>\lower1.06ex\hbox{$\kern-1.07em{\sim}$}$}}}
\def\lsimeq
{\hbox{\raise0.5ex\hbox{$<\lower1.06ex\hbox{$\kern-1.07em{\sim}$}$}}}

\def\xmm{{\it XMM-Newton }}

\def\asca{{\it ASCA}}

\def\xmm{{\it XMM-Newton}}
\def\chandra{{\it Chandra}}
\def\suzaku{{\it Suzaku}}
\def\rxte{{\it RXTE}}

\def\apj{ApJ}
\def\mnras{MNRAS}
\def\aap{A\&A}
\def\aapr{A\&ARv}
\def\apjl{ApJ}
\def\actaa{AcA}
\def\apjs{ApJS}
\def\araa{ARA\&A}
\def\pasj{PASJ}
\def\pasp{PASP}
\def\nat{Nature}

\def\exo{EXO~0748-676}
\def\axj{AX~J1745.6-2901}

\def\xis{XIS}
\def\xis1{XIS1}
\def\xis2{XIS2}
\def\xis3{XIS3}

\title[] 
 {{The puzzling orbital period evolution of the LMXB AX~J1745.6-2901}}

 \author[G.\ Ponti et al. ]
 {Gabriele~Ponti$^{1}$, Kishalay~De$^{1,2}$, Teodoro~Mu\~noz-Darias$^{3,4}$, Luigi Stella$^{5}$ 
 and Kirpal~Nandra$^{1}$
\\
   $^1$ Max Planck Institute f{\"u}r Extraterrestriche Physik,  Giessenbachstrasse, D-85748, Garching, Germany\\
   $^2$ Indian Institute of Science. Bangalore - 560012, India.\\
   $^{3}$ Instituto de Astrof\'isica de Canarias, 38205 La Laguna, Tenerife, Spain\\
   $^{4}$ Departamento de astrof\'isica, Univ. de La Laguna, E-38206 La Laguna, Tenerife, Spain\\
   $^{5}$ INAF-Osservatorio Astronomico di Roma, via Frascati 33, I-00040 Monte Porzio Catone (RM), Italy\\
}
\pagerange{\pageref{firstpage}--\pageref{lastpage}}
\usepackage{times}
\begin{document}
\label{firstpage}
 \maketitle
\begin{abstract}
The orbital period evolution of X-ray binaries provides fundamental clues to 
understanding mechanisms of angular momentum loss from these systems. 
We present an X-ray eclipse timing analysis of the transient low mass X-ray binary \axj. 
This system shows full eclipses and thus is one of the few objects for which accurate 
orbital evolution studies using this method can be carried out. 
We report on \xmm\ and \asca\ observations 
covering 30 complete X-ray eclipses spanning an interval of
more than 20 years. We improve the determination of the orbital period to a relative 
precision of $2\times10^{-8}$, two orders of magnitudes better than previous estimates. 
We determine, for the first time, a highly significant rate of decrease of the orbital 
period $\dot{P}_{orb}=-4.03\pm0.32\times10^{-11}$~s/s. 
This is at least one order of magnitude larger than expected 
from conservative mass transfer and angular momentum losses due to gravitational 
waves and magnetic breaking, and might result from non-conservative mass transfer.
Imprinted on the long term evolution of the orbit, we observe highly significant 
eclipse leads-delays of $\sim10-20$~s, characterised by a clear state dependence 
in which, on average, eclipses occur earlier during the hard state. 
\end{abstract}


\section{Introduction}

Eclipse timing provides us with powerful fiducial marks to investigate the orbital 
period evolution in low mass X-ray binaries (LMXBs; e.g. Chou 2014). 
This observable is closely associated with angular momentum variations of the system. 
In fact, any physical mechanism modifying the system angular momentum will 
inevitably modify the binary orbital period. 
The orbit can evolve as a consequence of mass transfer between the two stars 
(e.g. due to Roche lobe overflow), 
gravitational wave radiation (Landau Lifschitz 1958; Paczynski 1967; Faulkner 1971; 
Verbunt 1993), magnetic breaking (Eggleton 1976; Verbunt \& Zwaan 1981; 
Rappaport et al. 1983), mass loss from the companion, accretion disc winds, jets 
or other types of outflows (e.g. 
Fender et al. 2004; Ponti et al. 2012), and tidal interactions between the  
components of the binary system (Lear et al. 1976). 

The theory of angular momentum evolution due to mass transfer via 
Roche lobe overflow and gravitational wave radiation is well established and 
leads to quantitative predictions (Landau Lifschitz 1958; 
Paczynski 1967; Faulkner 1971; Verbunt 1993). However, angular momentum 
losses due to emission of gravitational waves only dominate at short orbital 
period  ($P_{\rm orb}\lsimeq2$~hr), while other mechanisms, such as magnetic 
breaking, are invoked to explain the orbit evolution at longer periods. 
The theory at the heart of the magnetic breaking model is based on the 
same physical principles through which magnetic stellar winds are known 
to decelerate the rotation of low-mass stars (e.g. Kraft 1967; Skumanich 1972; 
Sonderblom 1983). In synchronised binaries (typically the case 
of LMXBs), the angular momentum loss from the companion star  
is continuously re-distributed to the angular momentum of the whole binary system, 
therefore affecting its orbital period and binary separation (e.g. Taurus 2001).
Large uncertainties are involved in the extrapolation of magnetic breaking 
from isolated low-mass stars to the case of synchronised companion stars 
in LMXBs (e.g. see discussion in Knigge et al. 2011). 
Indeed, despite the large allowed range of model parameters  (possibly 
translating into more than one order of magnitude on efficiency of angular 
momentum removal), magnetic breaking often fails to reproduce measured values 
of the orbital period derivative, at least in the case of conservative mass transfer
(Di Salvo et al. 2008; Hartman et al. 2008; 2009; Hu et al. 2008; Burderi et al. 2010; 
Jain et al. 2010; Gonzalez-Hernandez et al. 2012; 2014). 
Additional mechanisms, typically involving outflows, are generally necessary 
to explain these discrepancies (Di Salvo et al. 2008; Hartman et al. 2008; 2009;
Burderi et al. 2010). 

Up to now, there are only a handful of LMXBs showing 
full eclipses for which a determination of the orbital 
period evolution has been possible (e.g. EXO~0748-676 Wolff et al. 2009; 
MXB~1659-29 Watcher et al. 2000; XTE J1710-281 Jain et al. 2011; 
GRS~J1747-312 in't Zand et al. 2003; Her X-1 Staubert et al. 2009). 

\axj\ is a dipping and eclipsing neutron star LMXB, showing type I bursts, 
discovered in \asca\ data during the 1993-1994 outburst (Maeda et al. 1996). 
It is one of the brightest X-ray transients of the Galactic centre region, located 
only $\sim1.5$~arcmin away from the supermassive black hole at the Galactic 
centre (Maeda et al. 1996; Kennea \& Skinner 1996; Hyodo et al. 2009;
Deegenar et al. 2012; Ponti et al. 2015a,b). During outburst it reaches a
brightness of 1--30\% the Eddington luminosity, alternating the hard (power-law 
dominated) and soft (thermal) states typically seen in 'atol' sources  
(Ponti et al. 2015; Mu\~noz-Darias et al. 2014; Gladstone, Done \& 
Gierlinski 2007; see section \ref{state}). 

The high column density of absorbing material 
($N_H\sim2\times10^{23}$~cm$^{-2}$) and the study of the black body radius 
during type I X-ray bursts suggest that \axj\ is located at or beyond the Galactic 
centre (Maeda et al. 1996; Ponti et al. 2015b). 
The first determination of the orbital period was performed by 
folding eclipses detected in the 1994 \asca\ data, yielding 
$P_{orb}=8.356\pm0.008$ hr ($30082\pm29$~s; Maeda et al. 1996, 
see also Sakano et al. 2001). 
Using \suzaku\ and \chandra\ observations taken between July 
and September 2007, Hyodo et al. (2009) measured an orbital period 
of $P_{orb}=30063.74\pm0.14$~s, consistent with the 
previous result. Due to its transient nature, crowding and high extinction 
(preventing detection below $\sim3$~keV), the orbital evolution 
of \axj\ has not been measured yet. 

Here we study for the first time the evolution of the orbital period of \axj\ 
over an interval of more than 20 years by using data from 15 years of \xmm\ 
monitoring campaign of the Galactic centre as well as data from the \asca\ archive.

\section{Observations and data reduction}

\subsection{\xmm}

We analysed all \xmm\ (Jansen et al. 2001) observations of \axj\ over the period 
November 2001 - April 2015. In addition to the observations 
analysed in Ponti et al. (2015a), we include ten new \xmm\ observations, taken 
during outburst (see Table \ref{EcTimes} for a full list of \xmm\ observations 
considered in this work). The EPIC pn camera aboard \xmm\ occasionally 
experiences time jumps. We checked that all the positive and negative jumps 
are properly corrected by the EPIC pn reduction pipeline\footnote{We checked 
that all time jumps as well as the unverified deltas and premature increments 
(see \xmm\ User Handbook, Ebrero et al. 2015) are properly accounted for. 
During {\sc obsid} 0402430401, 7 frames (from 4318574 until 4318580) are 
repeated, however this does not produce any time shift. 
During {\sc obsid} 0505670101, two telemetry gaps of 3~s and one of 10~s are 
observed. Moreover, the quadrant with CCD 4 experienced a negative jump of 
3~s, which is corrected. None of these produce any time shift. 
During {\sc obsid} 0723410501, three gaps of 634.2, 253.2 and 187.8~s, 
most probably due to telemetry gaps, happened in all CCDs. We manually 
checked that these intervals are very close to multiple integers of the frame 
time and that adding or subtracting a one full second time jump would not 
be multiple integer anymore. Therefore also these gaps do not produce any 
time shift. }.  
This is further confirmed by the fact that similar results are obtained with 
the EPIC MOS camera that is not affected by these time jumps. 
The arrival times of all events were corrected to the Solar 
system barycentre, applying the {\sc barycen} task of {\sc sas}. 

We processed the data sets using the latest version (14.0.0) of the \xmm\ Science 
Analysis System ({\sc sas}) and applied
the most recent (July 2015) calibrations. We use all the EPIC (Struder et al. 2001) 
cameras. Having higher effective area, our prime instrument is the EPIC-pn 
(Turner et al. 2001), but we also analysed the EPIC-MOS, 
for a consistency check. All observations have the medium filter applied 
and are in Full Frame mode, therefore the EPIC-pn and EPIC-MOS light curves have 
minimum time resolutions of 73.4~ms and 2.6~s, respectively. 

Since the main focus of this work is to measure the ingress/egress times and 
durations of the eclipses, increasing the number of source photons would 
provide a more accurate determination of the location of the transition points. 
Therefore, even if photon pile-up significantly 
affects the brightest observations we decided to retain all source 
photons (e.g. by not removing photons from the central part of the point spread 
function as in Ponti et al. 2015b). Pile-up is expected to have a negligible 
effect on the determination of the eclipse timing. 
The source and background photons are extracted from circular regions with 
40~arcsec and 160~arcsec radii, respectively. The former centred on \axj, 
while the latter covers a region free from bright point-like sources and low 
levels of diffuse emission. 

For each observation we identified periods of enhanced particle-induced 
background activity by inspecting the 6-15 keV source and background 
light curves binned with 5~s resolution. 
We filtered out periods with a background level 25~\% brighter than the 
average source count rate during the observation. 
Type I bursts have been removed, if clearly detected in 1~s source light curves. 
We note that type I burst were not detected either at the start, during or at the 
end of any eclipse.

To establish the X-ray state of the source (see section \ref{state}), we used the hardness 
intensity diagram shown in Fig. 2 of Ponti et al. (2015b), based on fluxes in the 3-6 and 
6-10~keV bands. When present, pile-up can significantly affect these measurements. 
To mitigate this we removed (only for this specific analysis) the central 9.25~arcsec 
in the source extraction region, such as done by Ponti et al. (2014; 2015). 
Using the EPIC pn data with a time resolution of 73.4~ms we searched for the 
presence of pulsations during observation OBSID: 0402430301 and 0690441801 (the longest 
observations where the source is bright in the soft and hard state, respectively) 
but did not detect any significant pulsed signal.

\subsection{\asca}

To significantly extend the time interval probed by the \xmm\ pointings, we also examined 
\asca\ observations. Data were obtained from the {\sc heasarc} database online, 
and screened event files were used in our analysis. As reported by 
Maeda et al. (1996), the source was found in outburst in 1993 and 1994. 
However, no eclipses were found in the 1993 data due to the short duration 
of the observation (additionally the source flux was 
about 5 times fainter than in 1994). Maeda et al. (1996) reported the 
discovery of eclipses with a 8.4~hr 
periodicity in the 1994 data.
We thus analysed this dataset, extracting photons from a radius 
of 0.8~arcmin as done by Maeda et al. (1996) and combining data from the two GIS 
instruments, which have a higher time resolution than the SIS. We filtered photons 
so as to include only those in the 3 - 10 keV range. Photon arrival times were corrected 
to the Solar system barycenter using the {\sc timeconv} task of {\sc ftools}. 
Finally, we found one eclipse fully covered by the observation, in the constructed 
light curve, which we used for our timing analysis.

Sakano et al. (2001) also reported a detection of the source by ASCA in 1997. 
However, the source flux was significantly weaker in this observation (similar 
to the levels in 1993), and only a hint of an eclipse was present in 
their folded light curve. Therefore we did not include the 1997 data in our analysis. 

\section{Eclipse Timing}
\label{timing}

The determination of the eclipse times in all data sets was done using the Bayesian Block 
technique (Scargle et al. 2013; Ponti et al. 2015c). The technique allows us to detect statistically 
significant changes in count rates for time-tagged event data, along with a determination
of the corresponding times. We first selected the source photons from the cleaned event files
and then we applied the Bayesian Block code on the photon arrival times. We performed 
this on all the data sets. We typically ran the Bayesian Block code over an entire 
observation, except for observations with intense dipping activity and long exposures, where
we split the event files into smaller segments, each containing one orbital period so that 
the routine converges in a reasonable amount of time. 
The eclipse ingress and egress times are singled out without ambiguity by using this technique.

\begin{figure}
\includegraphics[width=0.49\textwidth,height=0.3\textwidth,angle=0]{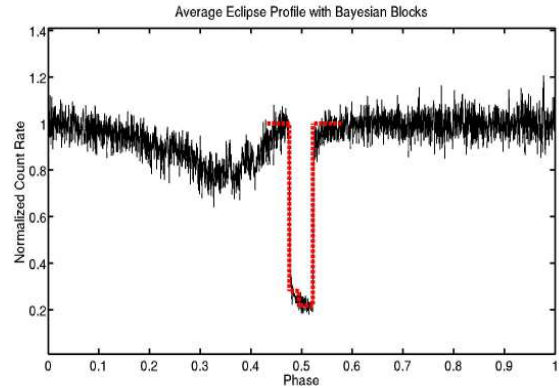}
\caption{Average, normalised count rate orbital profile in the 3-10~keV band
Bins are 10~s long. Individual profiles were aligned by using the eclipse 
centre time (which correspond to phase 0.5) as determined through the analysis 
described in the text.  
The red dotted line shows the results of the Bayesian blocks decomposition, 
at the time of the eclipse, determined during a bright soft state observation. 
After eclipse ingress a slow residual decay of the flux is observed. 
The average profile shows that the dipping phenomenon is more intense around phase 
0.1-0.5, inducing a significant trough (up to $\sim30$\%) of the average source flux, 
with a minimum around phase 0.3. 
Dipping events are sometimes observed just before the 
start of the eclipse, making the determination of the ingress time less reliable 
than that of egress time in some orbital cycles. }
\label{lcurve}
\end{figure}

Figure \ref{lcurve} shows the average source light curve during one orbit (obtained by
averaging over all orbits, after removing type I bursts); phases are defined so that 
the centre of the eclipse is at 0.5. 
At the time of eclipse ingress, the source flux abruptly decreases and then it 
sharply increases again by a similar amount at the time of the egress. 
Note that during the eclipse, a residual flux is present which displays 
a gradual decay near ingress.  
It has been suggested that this behaviour is due to a dust scattering halo (Hyodo et al. 2009). 
In agreement with this interpretation, we observe that the amplitude of this effect steeply 
decreases at higher energies and is of the same order of magnitude as expected 
for the large column density of material toward the source 
($N_H\sim2\times10^{23}$~cm$^{-2}$; Ponti et al. 2015b). 
A detailed study of this effect is beyond the scope of this paper. 

At low fluxes, most of the observed eclipses are well described by a single block. 
In these cases we define ingress and egress times as the start and end times of the block. 
However, for observations with higher source count rates, the presence of the slow decay 
during the eclipse induces one additional significant block at the start of the eclipse. 
Whenever an additional block appears, we define as ingress time, the start of the first 
block marking the eclipse and as egress time, the end of the last block marking 
the eclipse (see Fig. \ref{lcurve}). 
By using this method, we determined the start and end times of the 30 eclipses in the 
datasets that we analysed. 
Table \ref{EcTimes} gives the measured times using this technique.
\begin{table*}
\begin{center}
\begin{tabular}{ | c c c c c c c c }
\hline
Obs-ID & Cycle Number & PN-Count rate & Eclipse Ingress Time$\dag$ & Eclipse Egress Time & Error \\                  
       &              &   counts/sec   &       MJD        &          MJD        &  sec \\
\hline
\multicolumn{6}{c}{{\bf \asca}} \\
52005000  &  -21561  &  -  &  49612.31079  &  49612.32854  &  5.1 \\ 
\multicolumn{6}{c}{{\bf \xmm}} \\
0402430701  &  -8405  &  9.33  &  54190.05919  &  54190.07556  &  2.2  \\
0402430301  &  -8400  &  8.48  &  54191.79898  &  54191.81542  &  2.3 \\
0402430301  &  -8399  &  7.52  &  54192.14693  &  54192.16335  &  2.6 \\
0402430301  &  -8398  &  6.34  &  54192.49482  &  54192.51129  &  3.0 \\
0402430401  &  -8394  &  8.39  &  54193.88681  &  54193.90319  &  2.4 \\
0402430401  &  -8393  &  7.95  &  54194.23470  &  54194.25111  &  2.5 \\
0402430401  &  -8392  &  6.22  &  54194.58267  &  54194.59912  &  3.0 \\
0505670101  &  -7374  &  6.58  &  54548.80454  &  54548.82097  &  2.9 \\ 
0505670101  &  -7373  &  6.46  &  54549.15253  &  54549.16899  &  2.9 \\
0505670101  &  -7372  &  6.94  &  54549.50045  &  54549.51691  &  2.8 \\
0724210201  &  -1666  &  8.87  &  56534.95316*  &  56534.96964  &  2.2 \\
0724210201  &  -1665  &  10.54  &  56535.30115*  &  56535.31763  &  1.9 \\
0700980101  &  -1636  &  10.18  &  56545.39197  &  56545.40833  &  2.0 \\
0724210501  &  -1599  &  7.41  &  56558.26641  &  56558.28280  &  2.6 \\
0723410301  &  -1143  &  1.18  &  56716.93537  &  56716.95191  &  12.3 \\
0723410301  &  -1142  &  1.17  &  56717.28312  &  56717.29972  &  12.4 \\
0723410401  &  -1115  &  0.81  &  56726.67793  &  56726.69466  &  16.9 \\
0723410401  &  -1114  &  0.87  &  56727.02603  &  56727.04243  &  15.9 \\
0723410501  &  -1050  &  0.76  &  56749.29538  &  56749.31187  &  17.7 \\
0723410501  &  -1049  &  0.89  &  56749.64329  &  56749.65972  &  15.5 \\
0690441801  &  -1047  &  0.90  &  56750.33916  &  56750.35564  &  15.4 \\
0690441801  &  -1046  &  0.93  &  56750.68719  &  56750.70372  &  15.0 \\
0690441801  &  -1045  &  1.02  &  56751.03514  &  56751.05157  &  13.9 \\
0743630201  &  -617  &  1.26  &  56899.96128  &  56899.97786  &  11.7 \\
0743630301  &  -614  &  1.14  &  56901.00544  &  56901.02190  &  12.7 \\
0743630501  &  -533  &  1.18  &  56929.18989  &  56929.20639  &  12.3 \\
0743630601  &  -101  &  9.25  &  57079.50762*  &  57079.52486  &  2.2 \\
0743630801  &  -3  &  10.59  &  57113.60843  &  57113.62483  &  1.9 \\
0743630901  &  0  &  10.91  &  57114.65233  &  57114.66868  &  1.9 \\
\hline 
\end{tabular}
\caption{Measured eclipse times of the 29 \xmm\ eclipses and 1 \asca\ eclipse. 
The cycle number has been determined with respect to the reference time 
$T_0^e$ (see Section \ref{evolution}), which is considered to be cycle 0. 
$\dag$ Note that some of the ingress times (those marked with *) 
are affected by the dipping phenomenon, and thus may not be reliable (see 
discussion in Section \ref{timing}); we do not provide uncertainties 
for these measurements, but we expect them to be of the same order 
as those estimated for the egress times. 
} 
\label{EcTimes}
\end{center}
\end{table*} 

As a consistency check, we implemented the Bayesian Block technique on event files from 
the three EPIC cameras separately. Thus, for each eclipse we got three independent 
measurements of the eclipse times and their Bayesian Block uncertainties. 
A comparison between the uncertainties determined via the Bayesian Block 
approximation and those derived directly from the observed scatter (determined 
considering 12 measurements obtained by the EPIC cameras) suggests 
that the latter are more reliable (see Appendix 
and Fig. \ref{unc} for details) and were therefore used. 
In the case of the \asca\ eclipse, where such a scatter 
measurement was not possible, we used the uncertainty as inferred from the 
Bayesian Block algorithm.
For measuring the orbital evolution of the system, we used a weighted 
average\footnote{The weights have been taken as the average count rate 
of the source in the respective instrument during the orbital cycle under 
consideration.} of the eclipse times obtained from the three \xmm\ instruments 
(apart from the eclipse observed by \asca, where we directly used the measured 
eclipse time from the GIS instrument). 

On at least three occasions (cycle numbers -101, -1665 and -1666), we observe 
that the source was still dipping at the time of the eclipse ingress. 
Therefore, in order to proceed only with accurate measurements of the transition points, 
we used only the eclipse egress times to derive the orbital period evolution of the 
system. We note, however, that the analysis of the ingress times (after rejection of the  
cycles affected by dipping) provide results which are consistent within the errors of
those obtained from egress times. 

\section{Orbital evolution}
\label{evolution}

\begin{figure}
\includegraphics[width=0.49\textwidth,height=0.35\textwidth,angle=0]{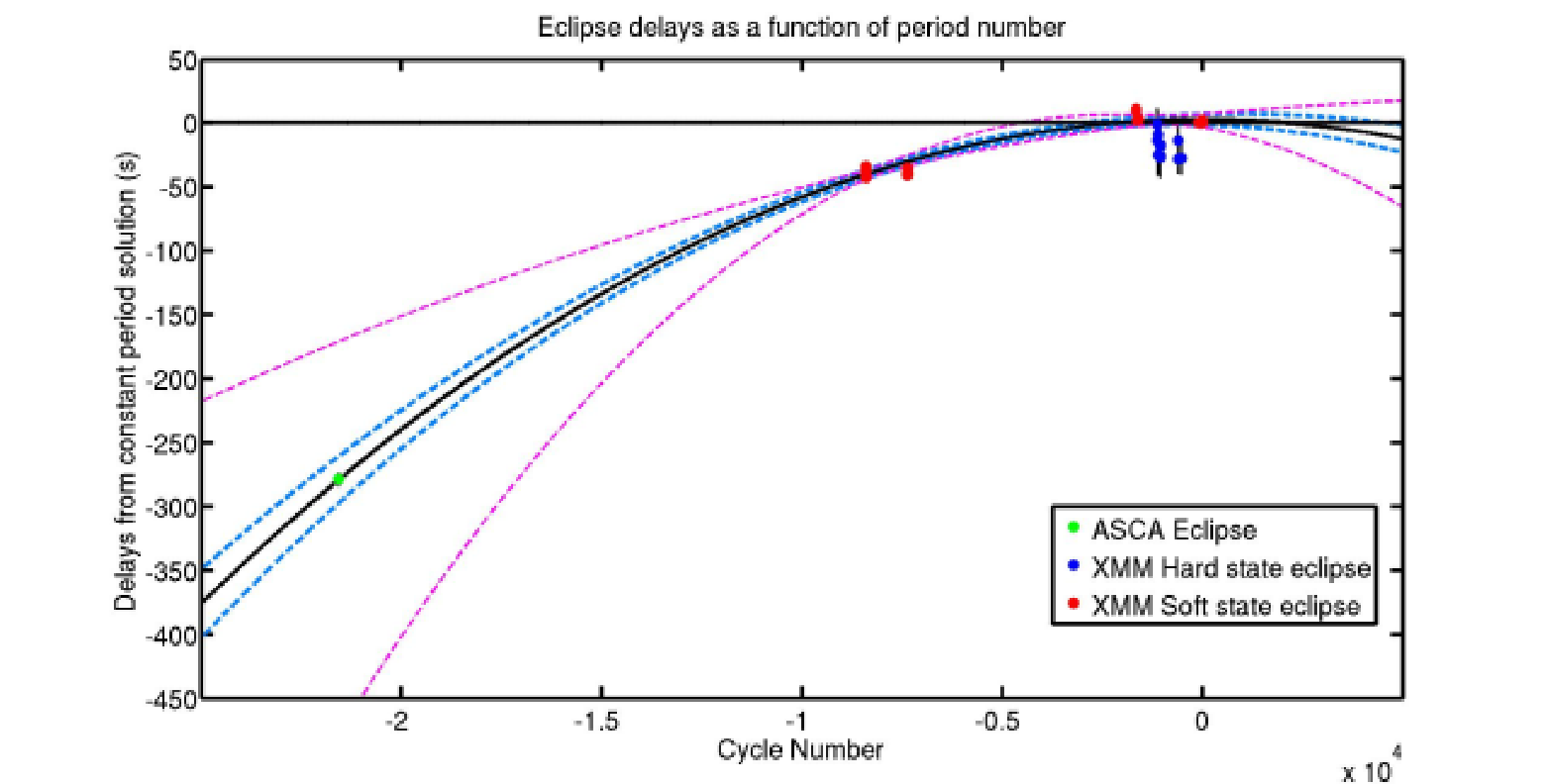}

\includegraphics[width=0.49\textwidth,height=0.35\textwidth,angle=0]{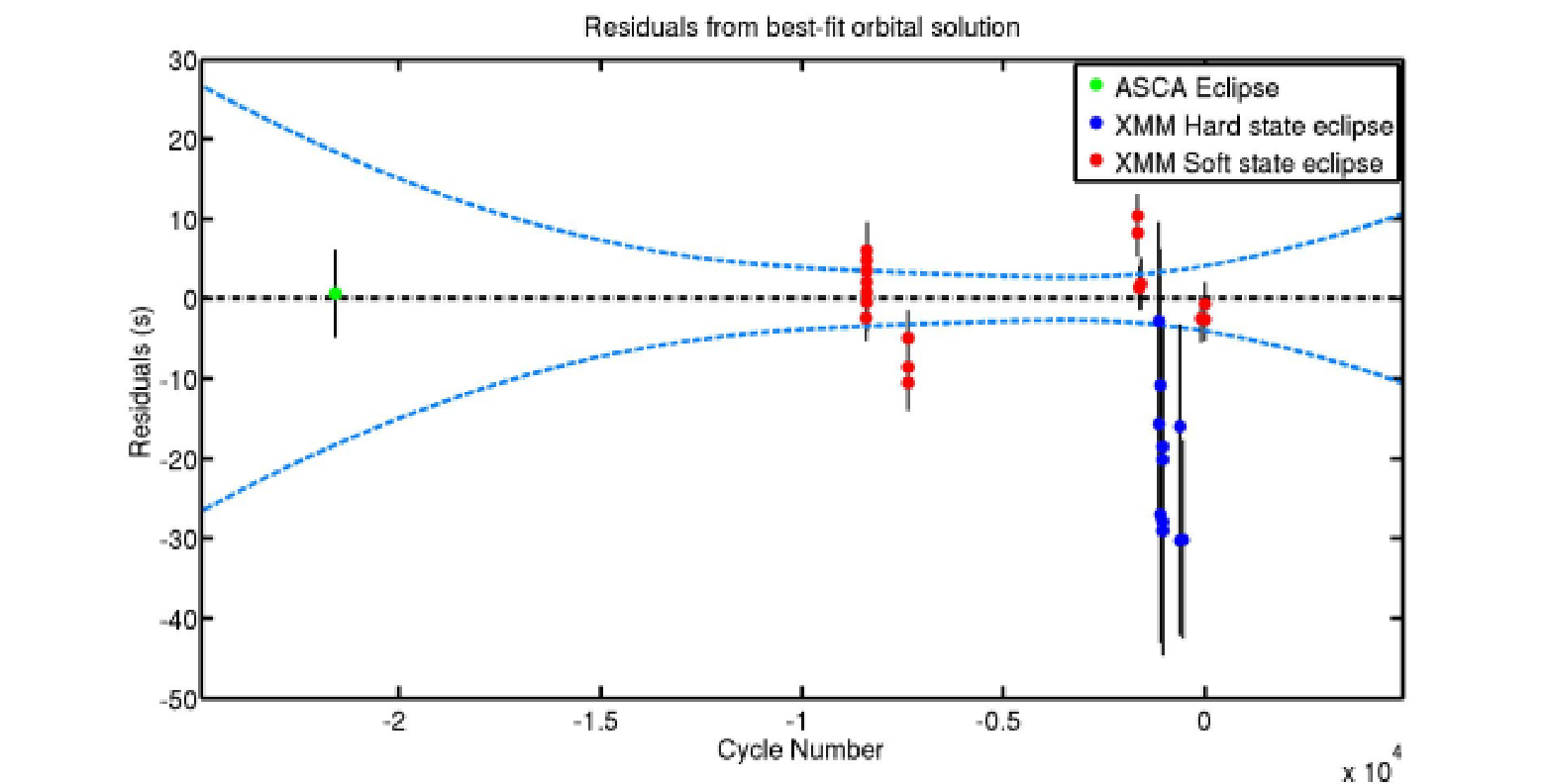}
\caption{{\it (Top panel)} 
Eclipse time delays (egress times) with respect to a constant 
orbital period model plotted versus the orbital cycle. 
$T_0$ (therefore cycle one) corresponds to MJD=57114.66871. 
The red and blue circles show the eclipse egress delays corresponding 
to soft and hard states of the source respectively, which were observed 
with \xmm\, whereas the green circle shows the eclipse egress 
delay derived from the single eclipse in the \asca\ data (1994).
The black solid and blue dashed lines show the best fit orbital solution of the 
combined \xmm\ and \asca\ data and the 68\% confidence region. 
The magenta dashed lines show the 68\% confidence uncertainty on the 
best fit orbital solution obtained fitting the \xmm\ data only. 
{\it (Bottom panel)} Residuals compared to the best fit orbital solution. 
Significant scatter is observed. }
\label{quad}
\end{figure}

The eclipse times determined using the method described above allows us to measure 
the evolution of the orbital period over a period of more than 20 years. We fitted the eclipse egress times with a curve of the form (consistent with a constant 
orbital period derivative model):
\begin{equation}
T_N^e - T_0^e = \delta T_0^e + P_{orb~0} N +\frac{1}{2} \dot{P}_{orb~0} P_{orb~0} N^2
\end{equation}
where $T_N^e$ is the eclipse egress time, $T_0^e$ is the reference eclipse egress time, 
$\delta T_0^e$ is the error in the determination of $T_0^e$, $P_{orb~0}$ is the orbital 
period at the epoch $T_0^e$, $\dot{P}_{orb~0}$ is the constant period derivative and 
$N$ is the (integer) number of orbital cycle counting from $T_0^e$. 
We choose $N$ to be the closest integer to $(T_N^e - T_0^e)/P_t$, where $P_t$ is 
the orbital period estimate given by Hyodo et al. (2008). In each case, we have 
verified that $|T_N^e-(T_0^e+N P_t)|<<P_t$ to ensure that the number of each orbital cycle 
is determined without ambiguity. 

We first test the possibility of a constant orbital period ($\dot{P}_{orb~0}=0$) through 
a linear fit. This yields an unacceptable solution with $\chi_{\nu}^2=29.1$, indicating 
that evolution of the orbital period is required by the data. 
We then used only the eclipse times obtained from \xmm\ observations to determine 
the ephemeris, and, in particular, to search for an orbital period derivative. 
We obtain: 
$P_{orb~0} = 30063.628 \pm 0.003$~s, and $\dot{P}_{orb~0} = ( -5.1 \pm 1.9 ) \times 10^{-11}$~s/s.
The reduced $\chi_{\nu}^2$ was found to be 3.30 for $\nu=25$. Thus, 
a non-zero orbital period derivative for the system was found by using the \xmm\ 
observations spanning an interval of $\sim8$ years (see magenta lines in Fig.~\ref{quad}). 
We then extended our baseline by more than 10 years by including now also the ASCA 
observation. The best fit parameters are (see also Tab. \ref{OrbPar}): 
$P_{orb~0} = 30063.6292 \pm 0.0006$~s, and 
$\dot{P}_{orb~0} = ( -4.03 \pm 0.27 ) \times 10^{-11}$~s/s.
We now obtain a reduced chi-squared value of $\chi_{\nu}^2=3.21$ with $\nu=26$. 
We find that the eclipse egress time obtained from the \asca\ observation, and the 
best-fit parameters using all eclipses, agree very closely with the values expected 
from the orbital solution obtained using solely  \xmm\ data (see Figure \ref{quad}). 
This suggests that this orbital solution has been valid for more than 20 years. 
We also test the possibility of a variable $\dot{P}_{orb~0}$, by fitting the eclipse 
times with a constant $\ddot{P}_{orb~0}$. This results in a $\ddot{P}_{orb~0}$ value 
consistent with $0$, and a reduced chi-square value of $\chi_{\nu}^2=3.22$, 
larger than that from the quadratic fit. 

The upper panel of Fig. \ref{quad} shows the eclipse time delays with respect 
to a constant orbital period model as a function of the orbital cycle. 
The solid black line shows the best fit solution with constant period derivative and 
the light blue dotted lines the corresponding one sigma uncertainty (the magenta 
dashed lines show the one sigma uncertainty for the \xmm\ data only). The orbital period 
of \axj\ clearly evolves in time, shortening by $\sim25$~ms and producing 
a delay of $\sim300$~s over the past twenty years. 

The bottom panel of Fig. \ref{quad} shows the residuals compared 
to the best fit model with a quadratic term. Significant residuals, as large 
as $10-20$~s, are observed (see \S~\ref{jitter}). 
In fact the best solution we obtained, with a $\chi_{\nu}^2=3.21$ for 
$\nu=26$, could be formally rejected because it does not reproduce 
all the scatter observed in the data. The presence of residual jitter 
with amplitude comparable to that observed here is typical of eclipsing LMXBs 
(e.g. Wolff et al. 2009; Jain et al. 2011). This is discussed further in the next section.

We investigated the reliability of the eclipse timings by applying 
a different technique to determine the eclipse ingress and egress time. 
We constructed a model light curve consisting of three blocks, corresponding 
to pre-eclipse, in-eclipse and post-eclipse regions, each with a constant 
count rate. We then used out a $\chi^{2}$ minimization routine to fit each of 
the eclipse light curves (binned to a time resolution of 5 seconds) to get the 
corresponding eclipse ingress and egress times for each eclipse. 
The start and end times of the in-eclipse region would then correspond to 
the eclipse ingress and egress times. We confirm that the eclipse timings, 
the orbital solution obtained and the residuals are consistent with that 
determined using the Bayesian Block analysis. 
\begin{table}
\begin{center}
\begin{tabular}{ | c c c c }
\hline
Parameter              &   Value                                & Units \\                  
\hline
$T_0^e$                & $57114.66871\pm0.00005$               & MJD \\                    
$P_{orb~0}$         & $30063.6292\pm0.0006$ & s  \\                          
$\dot{P}_{orb~0}$& $-4.03\pm0.27$                  & $10^{-11}$ s/s \\ 
\hline 
\end{tabular}
\caption{Best fit orbital solution for \axj\ derived from the analysis of the eclipse 
arrival times from 1994 to 2015. Errors are at 1 sigma confidence level on the last digit.
The value of $P_{orb~0}$ refers to $T_0^e$. } 
\label{OrbPar}
\end{center}
\end{table} 

\section{Residual jitter}
\label{jitter}

Both the residuals in Fig. \ref{quad} and the results of the best fit with 
a quadratic orbital solution, clearly shows the presence of residual jitter 
in the eclipse egress time with an amplitude as large as $\sim10-20$~s. 
We observe even higher residuals in the ingress time, about $\sim50-100$~s,
when dips are still in progress immediately before the eclipse onset
(clearly observed in at least three cases; see Tab.~\ref{EcTimes}). 
We interpret them as due to the dipping activity.
The smaller jitter in the eclipse egress time is most likely unrelated to dips, 
because: i) it is present also during orbits when no dips are detected; 
ii) dipping activity is very rarely observed shortly after the eclipse egress; 
iii) it is present also during orbits when no dips are detected. 
The \xmm\ calibrating team is excluding that the observed jitter has an 
instrumental origin. 

\subsection{State dependence}
\begin{figure}
\includegraphics[width=0.49\textwidth,height=0.34\textwidth,angle=0]{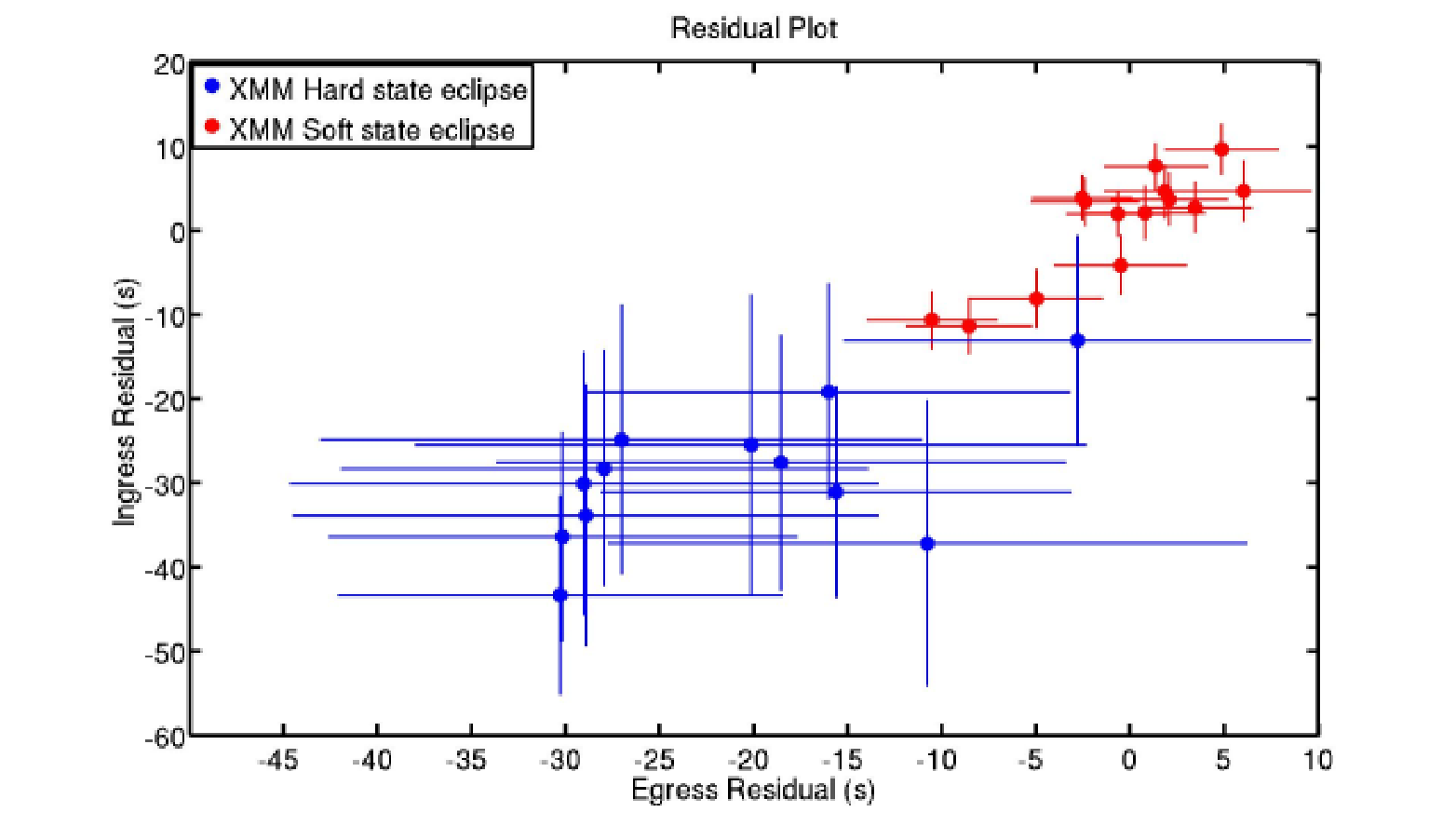}

\includegraphics[width=0.49\textwidth,height=0.34\textwidth,angle=0]{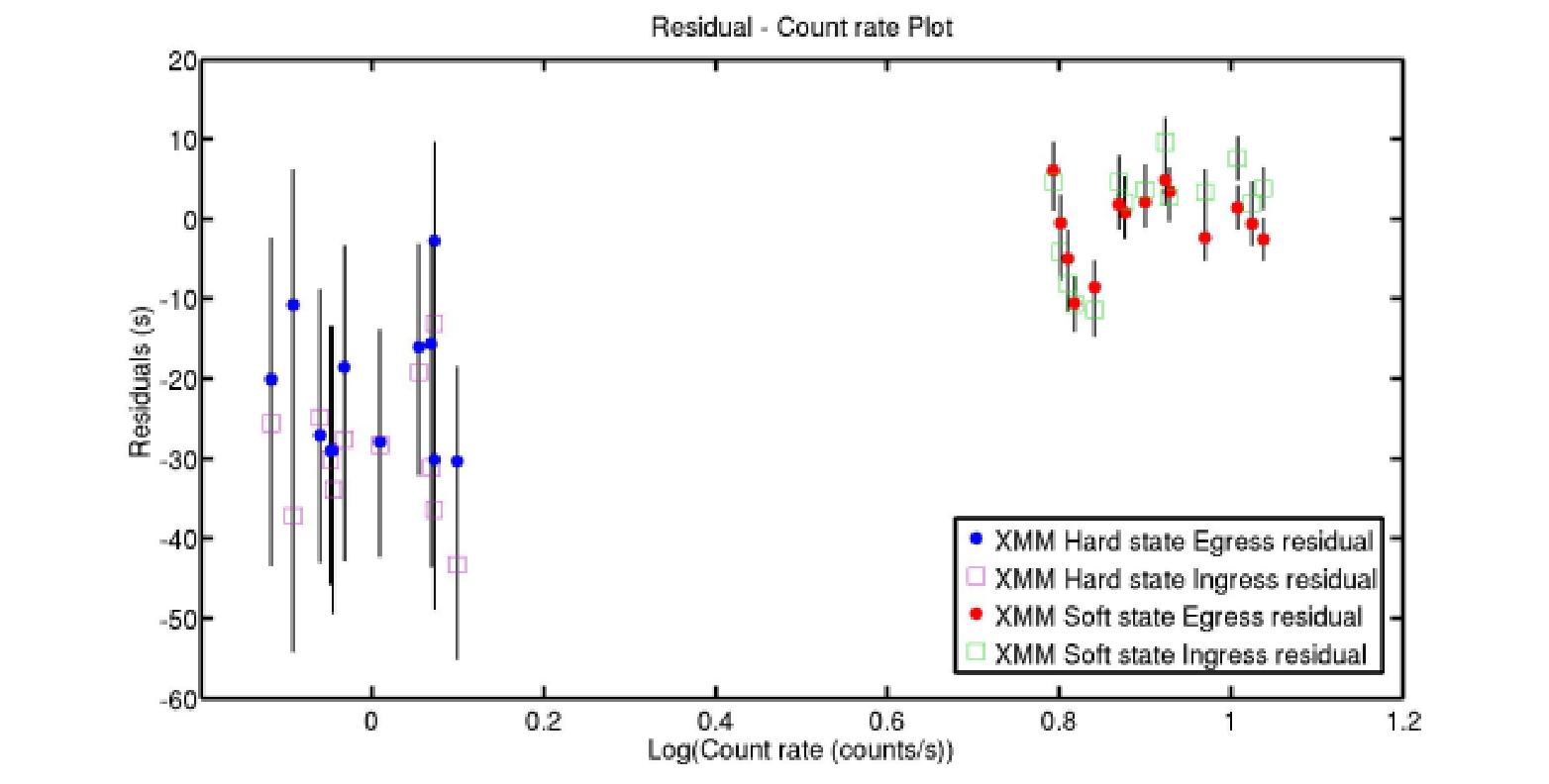}

\includegraphics[width=0.49\textwidth,height=0.34\textwidth,angle=0]{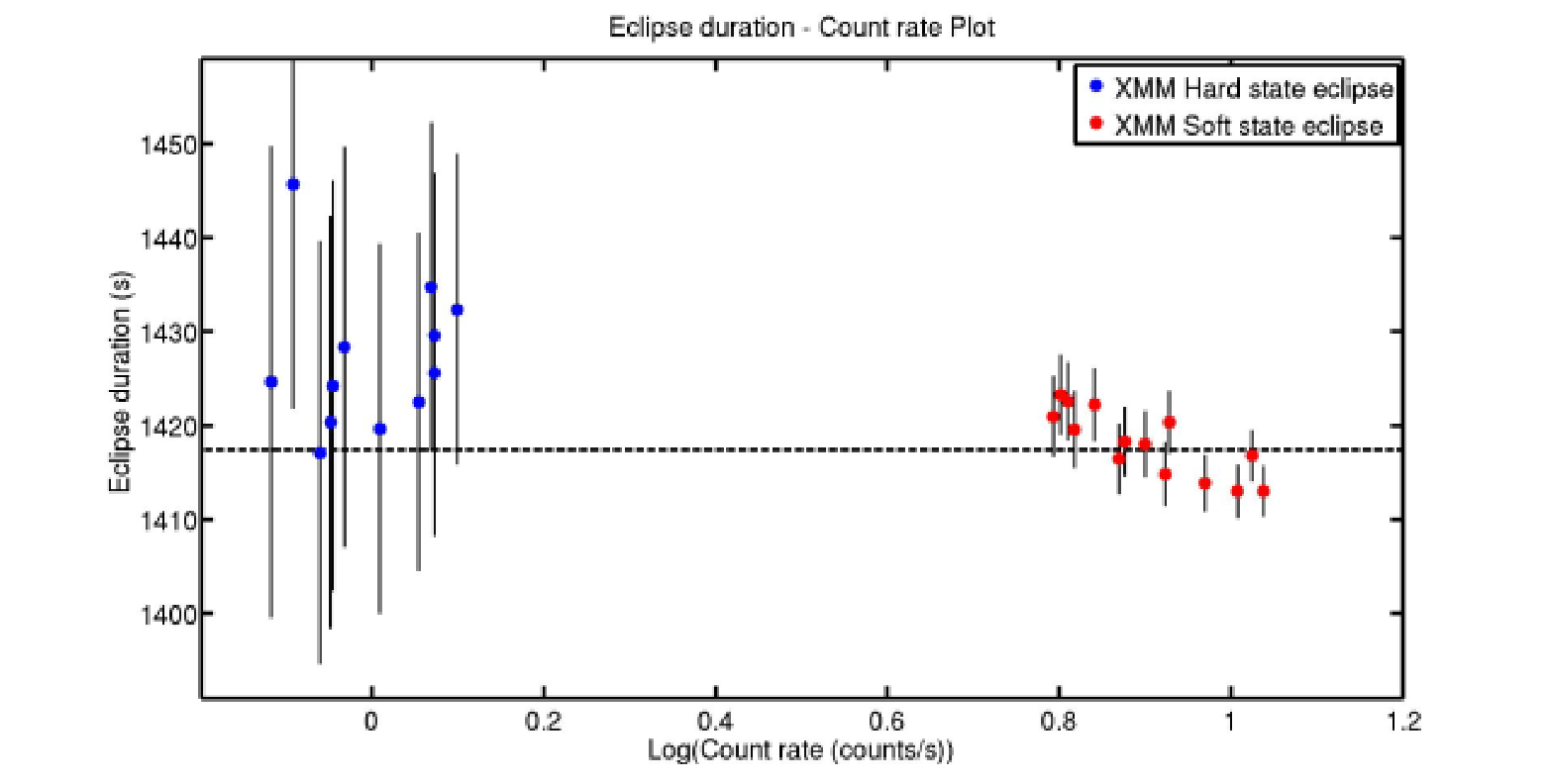}
\caption{ {\it (Top panel)} Ingress versus egress time residuals, compared to the best 
fit orbital solution shown in \S~\ref{OrbPar}. Shown here are the twenty 
six eclipses not significantly affected by the dipping phenomenon. 
The red points show the soft state observations, while the blue 
points indicate the hard state observations. The residuals in the ingress and egress 
time are well correlated. 
{\it (Middle panel)} Egress and ingress time residuals as a function of the source 
count rate with filled circles and open squares, respectively. 
Soft and hard state observations are shown in red and blue (green and magenta) 
for the egress (ingress) times, respectively. Ingress residuals show higher amplitude 
of variations. 
{\it (Lower panel)} Eclipse duration (defined as eclipse egress time minus eclipse 
ingress time) as a function of the observed count rate. The variation of the eclipse 
egress counterbalance the variation of the eclipse ingress, generating an almost 
constant duration of the eclipse. However, a small, but significant, shortening of the 
eclipse is observed at high fluxes. }
\label{res_state}
\end{figure}

The top panel of Fig.~\ref{res_state} shows the eclipse ingress as a function 
of eclipse egress residuals. Note that, in all panels of Fig.~\ref{res_state},
the blue (magenta) and red (green) points refer to the eclipses egress (ingress) time 
when the source was in the hard and soft state, respectively.
A clear correlation between these quantities is observed. 
Kendall's correlation coefficient is $r=0.91$, with a null hypothesis probability 
of $NHP\sim10^{-6}$.
The presence of such a correlation suggests that the jitter is generated 
by the eclipses being either delayed or advanced, while their duration remains
approximately constant. Indeed, as shown by the bottom panel of Fig. \ref{res_state}, a fit to 
the observed eclipse durations with a constant provides a reasonable fit 
($\chi_\nu^2=0.774$ for 24 dof; with an average duration of $\sim1419.1$~s, calculated 
for eclipses unaffected by dipping). However, we note a trend of decreasing 
eclipse duration with brightness in the soft state. Considering the duration and 
count rates during the soft state only, we do indeed observe a hint for a 
correlation (Kendall's correlation coefficient is $r=-0.65$, with a null hypothesis 
probability of $NHP\sim7\times10^{-4}$).
We also note that the eclipse ingress residuals show a higher scatter and 
span a larger range of values as compared to the egress residuals. 

The top panel of Fig.~\ref{res_state} shows a clear separation between the 
soft and hard state points. A Kolmogorov-Smirnov test of the distribution of 
soft and hard state delays (based on the eclipse egress times), indicates that the 
null hypothesis probability of the two residuals being extracted from 
the same population is $8\times10^{-7}$. The state dependence is such that 
the hard state eclipses in average occur at earlier times ($\sim 29$~s) 
than expected on the basis of the orbital solution. On the other hand, 
the eclipses happen roughly at the expected time (with an average 
delay of $\sim 0.69$s) when the source is in the soft state. 
This is not surprising because the source is typically brighter (having 
therefore smaller error bars) during the soft state. The middle and 
bottom panels of Fig.~\ref{res_state} shows that, indeed, the source 
is brighter in the soft state and the observed residuals are very well 
correlated with the source count rate (the Kendall's correlation coefficient 
is $r=0.5123$, with a null hypothesis probability $NHP\sim5\times10^{-5}$, 
corresponding to a significance of $\sim4$ sigma). 
We note that using the $\chi^2$ minimization technique to 
determine the eclipse timings also repoduces a clear state dependence of the 
residuals (KS nul	l hypothesis probability of $\sim10^{-4}$), with the 
residuals being correlated with the source count rate (Kendall's coefficient 
$0.567$, corresponding to a null hypothesis probability of $\sim10^{-5}$). 

\section{Discussion}

Fifteen years of \xmm\ monitoring of the Galactic centre joined with archival 
\asca\ data allowed us to time the eclipses and measure the 
evolution of the orbital parameters of \axj. We fitted the eclipse timing data 
with a  parabolic function, finding a solution valid for more than two decades. 
In particular, we determined the orbital period of \axj\ with a precision 
of 1 over $5\times10^{7}$, that is an improvement of over two orders of 
magnitudes compared to the previous best estimate (Hyodo et al. 2009). 
We also determine, for the first time, a highly significant derivative of the 
orbital period, which indicates that the system is shrinking and the 
orbital period decreasing at a rate of $-4.03\pm0.27\times10^{-11}$~s/s. 

We note that, in the best monitored eclipsing LMXB 
(\exo), eclipse timings obtained within different time intervals do produce 
inconsistent orbital solutions (Parmar et al. 1986; 1991; Asai et al. 1992; 
Corbet et al. 1994; Hertz et al. 1995; 1997; Wolff et al. 2002; 2009). 
In particular, detailed \rxte\ monitoring of the eclipse timing properties in 
\exo\ show that a unique orbital solution can be rejected at high significance 
and the orbital period evolution can be divided into at least three periods 
with abrupt changes between these (Wolff et al. 2002; 2009). 
Interestingly, other eclipsing systems have been suggested to undergo 
the same erratic behaviour (e.g. XTE~J1710-281; Jain et al. 2011). 
To check if this is the case for \axj, we fitted first the \xmm\ data alone, 
and then extrapolated the best fit orbital solution obtained within 
the period from 2007 to 2015, to the \asca\ data accumulated $\sim12$~years before 
(see \S~\ref{evolution}). We observe that the same orbital solution fitting 
the \xmm\ data, with a constant orbital period derivative, intercept the 
\asca\ data (see the magenta dashed lines in Fig.~\ref{quad}). 
This suggests the presence of a unique orbital solution in \axj\ and that 
the physical mechanism changing the angular momentum in \axj\ 
is most probably different from that at work in \exo. 

\subsection{Constraints on companion star}

By combining the Paczynski (1971) approximate formula for the Roche-lobe 
radius with Kepler's third law, it is possible to derive the average density of the 
companion star from the system orbital period. Assuming that the companion 
is a main sequence star, this would correspond to K0V companion of 
$M_2\simeq0.8$~M$_{\odot}$ (indeed such a star would fill the Roche lobe). 
As often observed in LMXBs, the companion star might be 
slightly evolved, implying $M_2\lsimeq0.8$~M$_{\odot}$.
Therefore, the mass ratio is constrained to $q=M_2/M_1\lsimeq0.57$ for 
$M_{NS}\geq 1.4 M_{\odot}$ 

A low mass main sequence star, loosing mass on a timescale much longer than 
the timescale on which the thermal equilibrium is established, is expected to 
have a mass-radius index $\zeta\simeq0.8$ ($R\simeq M^{0.8}$). 
When the mass loss is fast (adiabatic), the effective mass-radius index is 
$\zeta\simeq -1/3$ (Rappaport et al. 1982). 

\subsection{Conservative mass transfer} 

Conservative mass transfer induced by emission of gravitational waves 
and magnetic breaking can be approximated by 
\begin{eqnarray}
\dot{P}_{orb} = -1.4\times10^{-14}  m_1 m_{2,0.1} m^{-1/3} P^{-5/3}_{8hr} \times [1.0 + T_{MB}] \\
\times [(n-1/3) / (n+5/3 - 0.2m_{2,0.1}m_1^{-1} )]~s~s^{-1}
\end{eqnarray}
(see Burderi et al. 2010; Di Salvo et al. 2008; Verbunt 1993; Rapport et al. 1987), 
where $P_{8hr}$ is the orbital period in units of 8~hr; $m_1$, $m_{2,0.1}$ and $m$
are the mass (in Solar masses) of the primary and  
secondary and binary system ($m_1+m_2$), 
respectively (the secondary is expressed in units of $0.1~M_{\odot}$); 
$n$ is the index of the mass-radius relation of the secondary $R_2\propto M_2^n$, 
that is assumed to be in the range $0.6-0.8$; 
$T_{MB}$ represents the effects of  magnetic breaking (Eggleton 1976; 
Verbant \& Zwaan 1981; Tauris 2001; Burderi et al. 2010). 

Following Burderi et al. (2010; see also Verbunt \& Zwaan 1981; King 1988;
Verbunt 1993; Tauris 2001) we express the magnetic breaking term as 
\begin{equation}
T_{MB} = 49.4 (f/k_{0.277})^{-2} m^{1/3}_{2,0.1} m^{-4/3}_1 P^2_{8hr} ,
\end{equation}
where $k_{0.277}$ is the radius of gyration of the star $k$ in units of 0.277 (we assume 
$k_{0.277}=1$); $f$ is a dimensionless parameter of order unity with preferred values of 
$f=0.73$ (Skumanich 1972) or $f=1.78$ (Smith 1979). We chose here the value 
that maximises the effect of magnetic breaking ($f=0.73$), 
which results in $T_{MB}=129$. 

We observe that in a conservative scenario the orbital period derivative induced 
by gravitational radiation alone would be $\dot{P}_{orb} = -1.6\times10^{-14}~s~s^{-1}$, 
while it would be $\dot{P}_{orb} = -5.1\times10^{-12}~s~s^{-1}$, if magnetic breaking 
were at work\footnote{We note that for a lower index of the mass radius relation 
(e.g. $n=0.6$ instead of $n=0.8$), the orbital period evolution is slower, 
$\dot{P}_{orb} = -3.4\times10^{-12}$~s~s$^{-1}$. In the same way, if the companion 
is partly degenerate and thus has a lower mass (e.g. $m_2=0.4$), a slower 
evolution would be expected, $\dot{P}_{orb} = -1.5\times10^{-12}$~s~s$^{-1}$.}. 
We note that even the latter estimate is about one order of magnitude 
smaller than the observed value. We thus conclude that a conservative mass 
transfer scenario does not reproduce the behaviour of the source. 
Therefore, non conservative mass transfer appears to be required. 

\subsection{Connection with accretion state}
\label{state}
Superposed on the long-term evolution of the orbital period, highly significant 
jitter ($\sim10-20$~s) is observed. It is unlikely that the residual jitter originates from 
geometry variations of either the central X-ray source or the inner accretion 
disc, since these are small and get rapidly eclipsed by the rim of the companion star.
In fact, at a velocity of $\sim230$~km~s$^{-1}$, the companion star would swipe 
through the region where the bulk of the X-ray luminosity is produced, say 
$\sim100$~r$_g$ in less than $\sim1.2$~s (here r$_g$ is the gravitational radius 
defined as $r_g=GM/c^2$, where $c$ is the speed of light, $G$ is the gravitational 
constant and $M$ is the mass of the compact source). 

It is in theory possible that part of the observed jitter is associated to 
oscillations of the companion star atmosphere. 
For example, tidal forces could induce oscillations of the companion star. 
It is also well known that the magnetic activity in the Sun generates coronal 
loops extending several thousands kilometers above the photosphere.
The presence of such structure on the surface of the companion star 
would explain the amplitude of the observed jitter. One such event has 
been claimed in \exo\ (Wolff et al. 2007). 
However, were such effects at work in \axj, they would either modify 
the ingress and egress time in a random fashion (in the case of coronal loops), 
or in an anti-correlated way (i.e., ingress delayed and egress advanced or 
vice-versa, in case of oscillations of the companion star atmosphere). 
Therefore, in all these cases, a variation of the duration of the eclipse, 
with comparable amplitude to the observed jitter, would be expected. 
On the contrary we observe a nearly constant eclipse duration in \axj. 

The correlation between ingress and egress residuals suggest 
that jitter is caused by the whole companion star being either "delayed" 
or "advanced" with respect to the expected long term orbital period evolution. 
In particular, the hard state eclipses arrive, on average, $\sim29$~s 
before the expected time, while the soft state eclipses arrive 
very close to the expected time with an average delay of $\sim0.69$~s. 
A possibility would be the displacement of the center of mass 
of the system, by the presence of a third body. To investigate this, 
we fitted Doppler shifts in the time of arrival due to the orbital velocity 
in the presence of a third body in an elliptical orbit (adapting equations 
from Schreirer et al. 1972 and Mukherjee et al. 2006). 
The sparse sampling of the orbital evolution 
allows several possible solutions. However none of these solution is 
completely satisfactory. Indeed, they either provide an unlikely high 
eccentricity (higher than 0.999) or they predict (unlikely) large 
residuals ($\Delta t > 60$~s) during the unsampled periods. 
Moreover, the observed periodicities are multiples of half a year and 
likely arise from the \xmm\ observing window of the Galactic 
centre (observable for $\sim1.5$~months with a cadence of six months): 
therefore they might be spurious. Moreover, the connection between the 
eclipse timing and the X-ray 
source accretion state remains to be explained in the third body interpretation. 
We can also exclude that the center of mass could be modified by an 
asymmetric disc with a state dependent geometry, even though the outer 
disc bulge, generating dips, is observed to have a state-dependent behaviour. 
This would require an unrealistically high amount of mass 
($\sim10^{-3}$~M$_{\odot}$) to be stored in the bulge to produce the 
observed jitter.

We note that LMXB typically show durations of the eclipse 
transitions ({\it i.e.} duration of ingress and egress) of the order of 5-20~s 
(Cominsky \& Wood 1989; Wachter et al. 2000; Wolff et al. 2009).
Interestingly, these are comparable to the jitter observed in \axj.
Transition time scales are related to the atmospheric scale height 
of the companion star (indeed they correspond to $\sim1-4\times10^3$~km 
at a velocity of $\sim250$~km~s$^{-1}$). \axj\ shows the presence 
of a highly ionised plasma above the disc, with turbulent velocities of the 
order of $\sim500$~km~s$^{-1}$, temperatures of $kT\sim(4-10)\times10^6$~K 
and densities of order  $10^{12-13}$~cm$^{-3}$ (Ponti et al. 2012; 
2014; 2015). If such plasma travels to the orbit of the companion star 
without gaining additional angular momentum ({\it i.e.} maintaining the angular 
momentum characteristic of the outer parts of the accretion disc) it will 
not be in corotation with the companion star, which orbits at a speed of 
$\sim130$~km~s$^{-1}$. If so, this plasma will exert a ram pressure 
on the atmosphere of the star of the order of: 
$P_{ram}\sim4\times10^2 (n/10^{13}$~cm$^{-3}) (\Delta v/50$~km~s$^{-1})^2$~dy~cm$^{-2}$, 
where $n$ is the plasma density and $\Delta v$ the velocity difference 
(assumed to be $10^{13}$~cm$^{-3}$ and $50$~km~s$^{-1}$, respectively). 
This pressure is comparable to the pressure in the upper layers of a K type star 
atmosphere. Therefore, it could displace the position of the companion star 
atmosphere delaying both the ingress time and the egress time (although 
to a lesser extent, compared to the ingress time). 
The eclipse might appear therefore delayed and shortened during the soft state, 
with the ingress time being more affected than the egress time. 
No such effect would be present during the hard state, when the disc ionised 
atmosphere is not present (Ponti et al. 2015). 
We note that, in this scenario, during the soft state, the ram pressure would 
act on the companion star, slowing down its motion, and therefore affecting
the orbital evolution of the system. This would correspond to an additional 
contribution to the orbital period evolution of 
$\dot{P}\sim-1.5\times10^{-11}$~s/s (during the soft state only).

Finally we note that both the source luminosity and spectral energy 
distribution change significantly between the two canonical states. 
In particular, the bottom panel of Fig.~\ref{res_state} shows a correlation 
between the source count rate and the observed delay. 
Different levels of irradiation might affect the atmospheric layers of the 
companion star enhancing mass transfer rate and/or mass loss (e.g., Tavani 1991). 

Moreover, the geometry and physical conditions of the accretion disc 
are known to be significantly different in the two states. The hard state being 
characterised by a jet component and an inefficient disc, while the soft 
state showcasing a standard disc and an ionised atmosphere (Fender et al. 2004;
Done et al. 2007; Ponti et al. 2012). It appears plausible that different types of outflows, 
potentially driving away significant amounts of angular momentum and kinetic 
power, are present in the different states (e.g. jets in the hard state, winds 
in the soft state; Ponti et al. 2012; Fender \& Mu\~noz-Darias 2015). 
It is possible that either through irradiation, outflows or other mechanisms 
the system might lose angular momentum in a different way along the different 
accretion states. 

\appendix

\section{Uncertainties on block change times}
 
\begin{figure}
\includegraphics[width=0.49\textwidth,height=0.35\textwidth,angle=0]{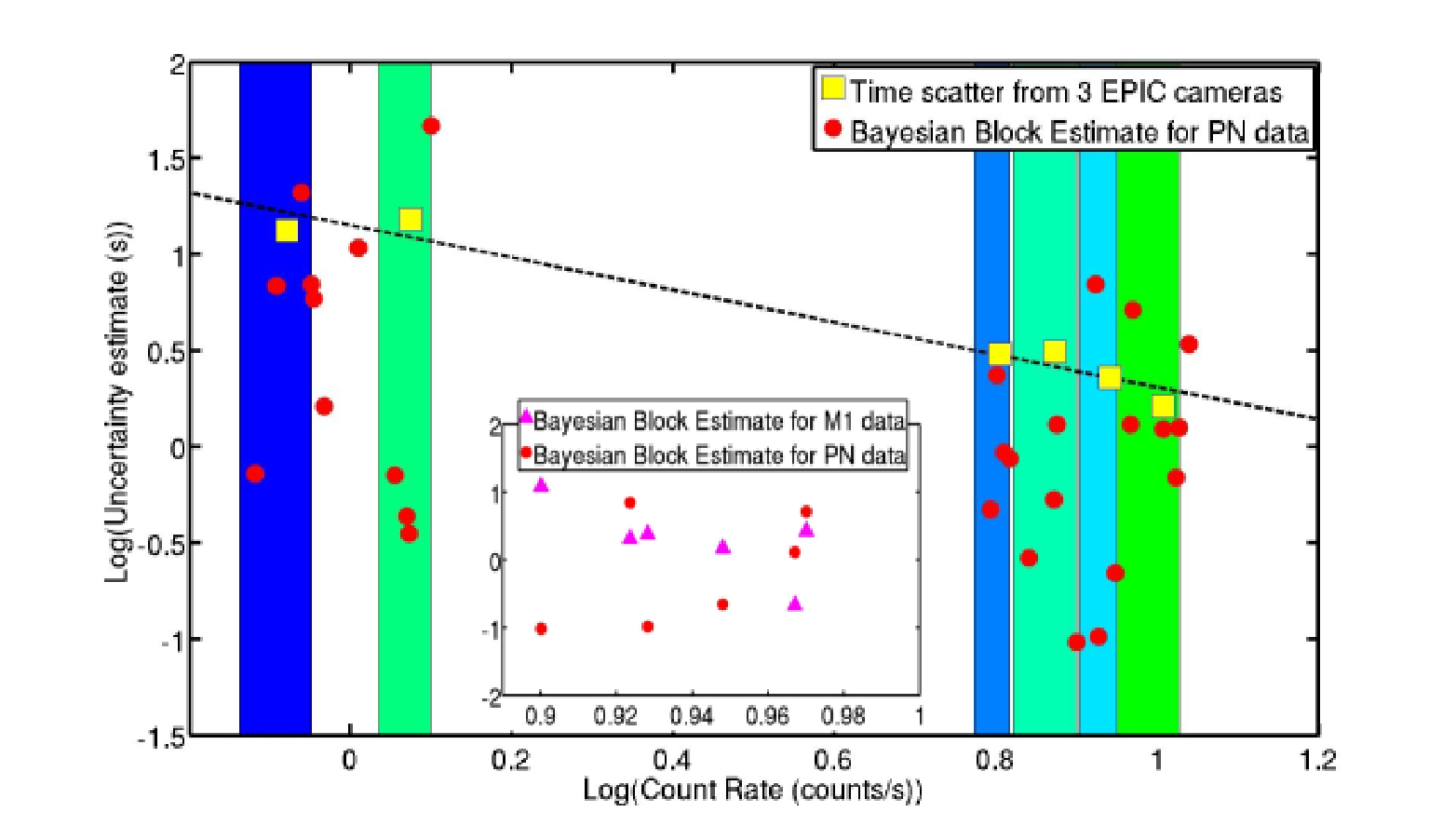}
\caption{{\it (Main panel)} The red circles show the uncertainties on the eclipse egress change times, 
as observed with the EPIC pn camera, determined through the Bayesian method described 
in Scargle et al. (2013). The yellow squares show the uncertainties derived from 
the scatter between the eclipse egress times measured by the different EPIC cameras 
averaged over orbits with similar sources flux. The coloured vertical regions indicate 
the flux intervals over which the average is performed. The black dashed line shows 
the best fit relation to the yellow points. 
{\it (Inset)} The red circles and magenta triangles show the uncertainties determined 
through the Bayesian block method for the EPIC pn and MOS1 cameras, respectively, 
over a narrow range of observed fluxes. The axes report the same quantities as 
the main panel. } 
\label{unc}
\end{figure}
The uncertainty on the block change times (i.e. the eclipses) was first determined using 
the method suggested in Scargle et al. (2013). This method allows us to compute 
the probability of change times for each block as a function of the photon arrival times, 
which peaks at the best fit block time. 
Since the probabilities thus obtained is a discrete distribution, we construct confidence 
intervals for change times by calculating the smallest symmetric interval around the best fit 
time which, in total, contains a given confidence level. The reported uncertainties, 
in all cases, are at a confidence level of 68\%. 
The red circles in Fig. \ref{unc} (both in the main figure and in the inset) show these 
uncertainties obtained from the EPIC pn data. The magenta triangles in the inset of 
Fig.~\ref{unc} show, instead, the Bayesian block estimates for the MOS1 data. 
We observe that, for similar source brightness, the uncertaintie determined in this way 
span a range from $0.1-7$~s, that appears relatively large. In particular, unreliably 
small uncertainties, as small as $\sim0.1-0.3$~s, are often observed. 
We note that, as clearly stated in Scargle et al. (2013), this method to determine 
the uncertainties on the block change times is approximate. 

We also estimated the uncertainties based on the observed scatter on the 
simultaneous observation of the eclipse change points (e.g. egress time) with the 
three EPIC cameras. The observed scatter (difference between the three 
measurements of the change times) will be equal or larger than the intrinsic 
uncertainty. 
For each eclipse we define as ingress and egress time the weighted average of the 
measurements obtained by the three EPIC cameras and we measure the scatter 
compared to this average. We then sort the eclipses in source mean flux (over 
the orbital period) and group the eclipses in order to have at least 4 eclipses 
in each bin. We finally average the observed scatter over the 4 eclipses 
(therefore over a total of 12 data points). 
The yellow squares in Fig. \ref{unc} show the averaged uncertainties obtained 
in this way. The vertical coloured stripes show the width of the bin over which 
the scatter has been averaged. 

We observe an overall good agreement between the two estimates of the uncertainties. 
However, we note that the uncertainties derived through the method suggested 
by Scargle et al. (2013) show a too large scatter. In particular, the inset of Fig.~\ref{unc} 
reports a comparison between the uncertainty measured by the Bayesian block method 
from the EPIC pn and MOS1 data (the axes report the same quantities as the main panel). 
We observe that in several occasions the MOS1 camera has uncertainties about one order 
of magnitude smaller than the pn camera, for the same eclipse, despite the smaller 
effective area. Therefore, we prefer to be conservative and use the second estimate 
of the uncertainty. We also note that, as expected, the uncertainties become smaller 
when the source is brighter. We then fit this trend, so that we can associate to any 
observed source flux an uncertainty on the determination of the change point and 
use the latter as uncertainties. 

\section*{Acknowledgments}

The authors wish to thank Michael Freyberg, Frank Haberl, Matteo 
Guainazzi and the \xmm\ calibration team for checking that the observed jitter 
has no instrumental origin. We thank Jan-Uwe Ness, Ignacio de la Calle 
and the rest of the \xmm\ scheduling team for the enormous support that 
made the new \xmm\ observations possible. We also thank Jonathan Ferreira, 
Pierre-Olivier Petrucci and Henri Gilles for very useful discussion. 
This research has made use primarily of data obtained with \xmm, an ESA 
science mission with instruments and contributions directly funded by ESA Member 
States and NASA, and on data obtained from the \chandra\ and \asca\ Data Archives.
The GC \xmm\ monitoring project is supported by the Bundesministerium 
f\"{u}r Wirtschaft und Technologie/Deutsches Zentrum f\"{u}r Luft- und Raumfahrt 
(BMWI/DLR, FKZ 50 OR 1408) and the Max Planck Society. KD would like to thank MPE for the
hospitality during his stay, and the German Academic Exchange Service (DAAD) for the
fellowship which supported his participation in the project.
TMD acknowledges support by the Spanish Ministry of Economy and Competitiveness 
(MINECO) under the grant AYA2013-42627.

\end{document}